\documentclass[epj]{svjour}
\usepackage{graphics}
\begin{document}
\title{Heavy quarkonium 2S states in light-front quark model}
\author{Tao Peng\inst{1} \and Bo-Qiang Ma\inst{1,2}
\thanks{\emph{email:} mabq@pku.edu.cn}%
}                     
%
%
\institute{School of Physics and State Key Laboratory of Nuclear
Physics and Technology, Peking University, Beijing 100871, China
\and Center for High Energy Physics, Peking University, Beijing
100871, China}
\date{Received: date / Revised version: date}
%
\abstract{ We study the charmonium 2S states $\psi'$ and $\eta_c'$,
and the bottomonium 2S states $\Upsilon'$ and $\eta_b'$, using the
light-front quark model and the 2S state wave function of harmonic
oscillator as the approximation of the 2S quarkonium wave function.
The decay constants, transition form factors and masses of these
mesons are calculated and compared with experimental data.
Predictions of quantities such as Br$(\psi' \to \gamma \eta_c')$ are
made. The 2S wave function may help us learn more about the
structure of these heavy quarkonia.
} 
\maketitle
%

\section{\label{sec:level1}Introduction}

Charmonium physics has long been an interesting issue as it is
related with both the perturbative and non-perturbative QCD.
Charmonia not only provide us with the opportunity to investigate
the interactions between the constituent quarks and the structure of
quarkonia, but also the chance to learn and understand the QCD
dynamics better. As exited states of charmonia, $\psi'$ and
$\eta_c'$ have been studied by many authors. The decay widths of
$\psi' \to e^+ e^-$ and $\eta_c' \to 2\gamma$ were calculated with
both relativistic and QCD radiative corrections~\cite{Chao96} and
the result $\psi' \to e^+ e^-$ is in agreement with experimental
data. The nonrelativistic potential model~\cite{Appelquist75} and
the Godfrey and Isgur (GI) model~\cite{Godfrey85,Barnes05} have
achieved much success, but their predictions of the decay widths of
$\psi' \to\gamma \eta_c~(\gamma \eta_c')$ are larger than
experimental data. The lattice QCD result~\cite{Dudek06,Dudek09} of
$J/\psi \to \gamma \eta_c$ is consistent with experimental data, but
the result of $\psi' \to \gamma \eta_c'$ has too large
uncertainties. The intermediate meson loop contribution to the
decays $\psi' \to \gamma \eta_c~(\gamma \eta_c')$ was investigated
recently~\cite{Li08,Li11}, and the results are closer to
experimental data. We also investigate these decays, using
light-front formalism and the harmonic oscillator wave functions as
the approximate wave functions of the 1S and 2S quarkonia. In fact,
there are still some puzzles concerning $\psi'$, such as the
well-known ``$\rho \pi$
puzzle"~\cite{rhopipuzzle1,rhopipuzzle2,rhopipuzzle3,rhopipuzzle4},
and the recent unanticipated small experimental value of Br$(\psi'
\to \gamma \eta) $/Br$(\psi' \to \gamma
\eta')$~\cite{psiprimepuzzle1,psiprimepuzzle2}. For the ``$\rho \pi$
puzzle", Ref.~\cite{Brodsky97} suggested the explanation that the
$\psi' \rho \pi$ coupling is suppressed due to the mismatch between
the nodeless wave function of the $\bar{c}c$ in the
$|u\bar{d}\bar{c}c\rangle$ Fock state of $\rho$ and the one-node 2S
$\bar{c}c$ wave function of $\psi'$, and our postulation of the 2S
wave function of $\psi'$ may be able to offer a numerical
realization for this explanation. The BES and CLEO collaborations
have conducted many experimental measurements on 1S and 2S
charmonia, and the decay mode $\psi' \to \gamma \eta_c'$ is being
studied by BES-III. It is then important to learn carefully about
the structures and decay mechanisms of the 2S charmonia.

The 2S bottomonia have been studied by some experiments, but many
data about them are still not available, such as the mass and decay
data of $\eta_b'$~\cite{PDG10}. In Ref.~\cite{Chao96}, the decay
modes $\eta_b' \to 2\gamma$ and $\Upsilon' \to e^+ e^-$ were studied
by considering both relativistic and QCD radiative corrections, and
predictions were made. With the same method of studying the 2S
charmonia in the light-front quark model, we can study more decay
modes, and calculate the masses of these bottomonia in this paper.
From an experimental viewpoint, a large amount of bottomonia and
their excited states could be produced at the forthcoming LHC or by
the Belle experiment in the near future, and they could provide
important tests of different predictions.

Moreover, heavy quarkonia, especially charmonia and bottomonia, act
as improtant diagnostic tools to probe the properties of the
background QCD matter, such as of the formation of the quark-gluon
plasma (QGP), in heavy ion collisions at RHIC and LHC~\cite{Wu10}.
As has been pointed in Ref.~\cite{Wu10}, the study of
heavy-quarkonium suppression at RHIC energy, which might be a
signature for the QGP formation, calls for the knowledge of the
light-front wave functions of the quarkonia: $f(x_\perp,x_\perp^
\prime,\tau_0)=\varphi(x_\perp)\varphi^*(x_\perp^\prime)$, where
$\varphi(x_\perp)$ can be taken as the Fourier transform of our
light-front momentum space wave functions for the 1S or 2S quarkonia
(Eqs.~(\ref{wfk1S}) and (\ref{wfk2S})), and $f(x_\perp,
x_\perp^\prime,\tau_0)$ are the essential quantities to calculate
the transverse momentum distribution of quarks. Thus our wave
functions can not only help us understand the structure of heavy
quarkonia themselves, but also be used as inputs for other physical
studies.

This paper is organized as follows. In Sec.~\ref{sec:level2}, we
describe the light-front quark model and the 2S state wave function
for the quarkonia. In Sec.~\ref{sec:level3}, we present our
numerical results of the decay constants, form factors and masses of
these charmonia and bottomonia, and compare them with experimental
data. A brief summary is given in Sec.~\ref{sec:level4}.

\section{\label{sec:level2} Model description}

Heavy quarkonia have been studied by non-relativistic
treatments~\cite{Appelquist751,Appelquist752,Appelquist753}, but in
some occasions related to non-perturbative scales, they have to use
model dependent methods. And as the virtual photon momentum $Q^2$
increases, the relativistic effects become important. So it is
useful to study quarkonia in a relativistic treatment. Several
powerful non-perturbative tools have been developed to study the
structure and decays of mesons, such as the QCD sum-rule technique
and the lattice gauge theory. The light-front quark model is also an
important model to do such
studies~\cite{Brodsky821,Brodsky822,Brodsky98}, and it has a number
of salient features. Light-front quark model includes some important
relativistic effects that are neglected in the traditional
constituent quark model, and the vacuum in the light-cone coordinate
is simple because the Fock vacuum is the exact eigenstate of the
full Hamiltoian. Light-front quark model has been successfully
applied in many investigation of hadron
structures~\cite{Xiaoetal1,Xiaoetal2,Xiaoetal3,Xiaoetal4,Xiaoetal5,Ma93,MaMeson1,MaMeson2,MaNucleon1,MaNucleon2,MaNucleon3,MaNucleon4,MaNucleon5,MaNucleon6,MaNucleon7}.

In the light-front quark model, the states of quarkonia can be
described by the Fock state expansion
\begin{eqnarray}
|M\rangle &=& \sum |q\bar{q}\rangle \psi_{q\bar{q}}
        + \sum
        |q\bar{q}g\rangle \psi_{q\bar{q}g} + \cdots,
\end{eqnarray}
and to simplify the problem, we adopt the lowest order of the above
expansions and take only the quark-antiquark valence states of the
mesons into consideration.

The quarkonium wave function in light-front formalism
is~\cite{Brodsky821,Brodsky822,Huang94}
\begin{eqnarray}
|M (P^+, \mathbf{P}_\perp, S_z) \rangle
   &=&  \int \frac{\mathrm{d} x \mathrm{d}^2
        \mathbf{k}_{\perp}}{\sqrt{x(1-x)}16\pi^3}\nonumber\\
        &&\cdot\phi(x,\mathbf{k}_{\perp})
        \chi_M^{S_z}(x,\mathbf{k}_{\perp},\lambda_1,\lambda_2),
\end{eqnarray}
with the momentum of the struck quark being $(xP^+, [m^2+(x
\mathbf{P}_\perp+\mathbf{k}_{\perp})^2]/(xP^+), x
\mathbf{P}_\perp+\mathbf{k}_{\perp})$, and  $\lambda_i$ being the
helicity of the $i$-th constituent quark.
$\phi(x,\mathbf{k}_{\perp})$ is the radial wave function, and
$\chi_M^{S_z}(x,\mathbf{k}_{\perp}, \lambda_1, \lambda_2)$ is the
light-front spin wave function, which is related to the instant-form
spin wave function by the Melosh-Wigner rotation
~\cite{Kondratyuk80,Melosh741,Melosh742,Melosh743,MaNucleon1,MaNucleon2,MaNucleon3,MaNucleon4,MaNucleon5,MaNucleon6,MaNucleon7}
\begin{equation}
\left\{
\begin{array}{lll}
\chi_i^\uparrow(T) &=& w_i[(k_i^+ +m_i)\chi_i^\uparrow(F)-k_i^R
\chi_i^\downarrow(F)],\\
\chi_i^\downarrow(T)&=& w_i[(k_i^+ +m_i)\chi_i^\downarrow(F)+k_i^L
\chi_i^\uparrow(F)],
\end{array}
\right.
\end{equation}
where $w_i=1/\sqrt{2k_i^+ (k^0+m_i)}$, $k^{R,L}=k^1\pm k^2$,
$k^+=k^0+k^3=x \mathcal{M}$, $m_i$ is the mass of the constituent
quark, and the invariant mass of the composite system $\mathcal{M}
\equiv \sqrt{(\mathbf{k_\perp}^2+m_1^2)/x+(\mathbf{k_\perp}
^2+m_2^2)/(1-x)}$. The Melosh-Wigner rotation is an important
ingredient of light-front quark model and plays an essential role in
explaining the ``proton spin puzzle"
~\cite{MaNucleon1,MaNucleon2,MaNucleon3,MaNucleon4,MaNucleon5,MaNucleon6,MaNucleon7}.
In the above formalism, the Drell-Yan-West ($q^+=0$)
frame~\cite{dywframe1,dywframe2} is used because only valence
contributions are needed in this frame when studying the decay of
quarkonia.

For the radial wave function $\phi$, the harmonic oscillator wave
function has been adopted to describe the 1S state mesons
~\cite{Huang94,Ma93,MaMeson1,MaMeson2}, and it can well explain
experimental data. So we try to go further to use the 2S state
harmonic oscillator wave function as the approximate wave function
of the 2S quarkonia. The wave functions of the 1S and 2S states of
the non-relativistic 3-dimensional isotropical harmonic oscillator
in momentum space are
\begin{eqnarray}
\varphi^{1S}(\mathbf{p})&=&\frac{1}{\pi^{3/4}(\alpha \hbar)^{3/2}}
\exp(-\frac{p^2}{2\alpha^2 \hbar^2}),\label{wfp1S}\\
\varphi^{2S}(\mathbf{p})&=&\frac{\sqrt{6}}{3\pi^{3/4}(\alpha \hbar)
^{7/2}}(p^2-\frac{3}{2}\alpha^2 \hbar^2) \exp(-\frac{p^2}{2\alpha^2
\hbar^2}), \label{wfp2S}
\end{eqnarray}
where $\alpha=\sqrt{\mu \omega/\hbar}$, $\mu$ and $\omega$ are the
mass of the oscillating particle and the frequency of the
corresponding classical oscillator respectively.

We use the connection between the equal-time wave function in the
rest frame and the light-front wave function suggested by
Brodsky-Huang-Lepage~\cite{Brodsky821,Brodsky822,Huang94}, for the
quarkonia with $m_1=m_2\equiv m_q$,
\begin{equation}
p^2 \longleftrightarrow \frac{\mathbf{k}_{\perp}^2+m_q^2}
{4x(1-x)}-m_q^2 \label{connection}
\end{equation}
and we use the prescription in Ref.~\cite{Chung88} to extend the
non-relativistic form wave function into a relativistic
one~\cite{Ma93}. Then we have the corresponding relativistic wave
functions in light-front formalism
\begin{eqnarray}
\phi^{1S}(x_i,\mathbf{k}_{i \perp})&=&\frac{4\pi^{3/4}}{\beta^{3/2}}
\sqrt{\frac{\partial k_z}{\partial x}}
\exp(-\frac{\mathbf{k}^2}{2\beta^2}),\label{wfk1S}\\
\phi^{2S}(x_i,\mathbf{k}_{i \perp})&=&\frac{4\sqrt{6}\pi^{3/4}}
{3\beta^{7/2}}\sqrt{\frac{\partial k_z}{\partial x}}
(\mathbf{k}^2-\frac{3}{2}\beta^2)\exp(-\frac{\mathbf{k}^2}{2\beta^2}),
\label{wfk2S}
\end{eqnarray}
where $\beta$ is the parameter equivalent with $\alpha$ in
Eqs.~(\ref{wfp1S}) and (\ref{wfp2S}), and its value can be chosen to
fit experimental data. The longitudinal momentum $k_z=(x-1/2)
\mathcal{M}+(m^2_2-m^2_1)/2\mathcal{M}$, one can easily check that
this is equivalent to Eq.~(\ref{connection}). The factor
$\sqrt{\partial k_z/\partial x}$ in the above two equations comes
from the Jacobian of the transformation $(x, \mathbf{k}) \to
(\mathbf{k}, k_z)$, and the normalization factors are from the
requirement of the normalization of the total wave function
~\cite{Choi07}.

Using the above formalism and wave functions, we can calculate the
decay constants and transition form factors of the quarkonia
~\cite{Qian09,Choi97}.

In the $V\rightarrow e^+e^-$ process, the decay constant of the
vector meson $V$ is defined by
\begin{equation}
\langle 0| j_\mu |V(p,S_z)\rangle = M_V f_V \epsilon_\mu(S_z),\\
\end{equation}
and with the same method as Ref.~\cite{Qian08}, we have, for the
vector quarkonium,
\begin{eqnarray}
f_V &=&2\sqrt{6}~e_q\int \frac{\mathrm{d}x
\mathrm{d}^2\mathbf{k}_\perp}{16\pi^3} \frac{1}{\sqrt{x(1-x)}}
~\phi_V(x,\mathbf{k}_\perp)\nonumber\\
&&\cdot\frac{2\mathbf{k}_\perp^2+m_q(\mathcal{M}+2m_q)}{\sqrt{\mathbf{k}
_\perp^2+m_q^2}(\mathcal{M}+2m_q)},
\end{eqnarray}
where $m_q$ and $e_q$ is the mass and electric charge of the
constituent quark of the quarkonium respectively~($e_q=2/3$ for
charmonia, and $-1/3$ for bottomonia).

In the $P\rightarrow\gamma\gamma^*$ process, the transition form
factor of the pseudoscalar meson $P$ is defined by
\begin{equation}
\langle \gamma(p-q)|J^\mu |P(p,\lambda)\rangle = ie^2F_{P\rightarrow
\gamma\gamma^*}(Q^2)
        \varepsilon^{\mu\nu\rho\sigma}p_\nu\epsilon_\rho(p-q,\lambda)
        q_\sigma,
\end{equation}
and we have the formula for the pseudoscalar quarkonium
\begin{eqnarray}
F_{P\rightarrow \gamma\gamma^*}(Q^2)&=&4\sqrt{6}~e^2_q \int
\frac{\mathrm{d}x \mathrm{d}^2\mathbf{k}_\perp}{16\pi^3}~
\phi_P(x,\mathbf{k}_\perp)\nonumber\\
&&\cdot\frac{m_q}{x\sqrt{\mathbf{k}_\perp^2+m^2}}
\frac{x(1-x)}{m_q^2+\mathbf{k}_\perp^{'2}},
\end{eqnarray}
where $\mathbf{k}'_\perp=\mathbf{k}_\perp-(1-x)\mathbf{q}_\perp$,
and $Q^2=-q^2=\mathbf{q}_\perp^2$, $q$ is the momentum of the
virtual photon.

The radiative transition form factor between a vector meson $V$  and
a pseudoscalar meson $P$ is  defined by
\begin{equation}
\langle P(p')|J^\mu |V(p,\lambda)\rangle = ieF_{V\rightarrow \gamma
P}(Q^2)
        \varepsilon^{\mu\nu\rho\sigma}\epsilon_\nu(p,\lambda)
        p'_\rho p_\sigma,
        \label{fvp}
\end{equation}
and we have
\begin{eqnarray}
F_{V\rightarrow \gamma P}(Q^2)=4~e_q\int \frac{\mathrm{d}x
\mathrm{d}^2\mathbf{k}_\perp}{16\pi^3}~\phi_P(x,\mathbf{k}_\perp')
\phi_V(x,\mathbf{k}_\perp)\nonumber\\
\cdot \frac{m_q(\mathcal{M}+2m_q)(1-x)+ 2(1-x)\mathbf{k}_\perp^2
\sin^2\theta} {(\mathcal{M}+2m_q)\sqrt{\mathbf{k}_\perp^2+m_q^2}
\sqrt{\mathbf{k}_\perp^{'2}+m_q^2}},
\end{eqnarray}
where $\theta$ is the angle between $\mathbf{k}_\perp$ and
$\mathbf{q}_\perp$.

The above quantities are related to the decay width of the
quarkonium by~\cite{Qian09}
\begin{eqnarray}
\Gamma(V\rightarrow e^+ e^-) &=& \frac{4\pi\alpha^2f_V^2}{3M_V},\\
\Gamma(P\rightarrow\gamma\gamma) &=& \frac{1}{4}\pi\alpha^2 M_P^3
|F_{P\rightarrow\gamma\gamma^*}(0)|^2, \label{decay width a}\\
\Gamma_{V\rightarrow \gamma P}
    &=&\frac{\alpha}{3}
    \left|F_{V\rightarrow \gamma^*P}(0)\right|^2
    \left(\frac{M_V^2-M_P^2}{2 M_V}\right)^3.
\label{decay width b}
\end{eqnarray}

We can also calculate the mass of the quarkonium, using the
QCD-motivated Hamiltonian for mesons~\cite{Choi99}
\begin{equation}
H_{q\bar{q}}=\sqrt{m^2_q+\mathbf{k}^2}+\sqrt{m^2_{\bar{q}}
+\mathbf{k}^2}+V_{q\bar{q}},
\end{equation}
where $\mathbf{k}$ is the momentum of the constituent quark, and
\begin{equation}
V_{q\bar{q}}=a+br^2-\frac{4\alpha_s}{3r}+\frac{2\mathbf{S}_{q}
\cdot\mathbf{S}_{\bar{q}}}{3m_q m_{\bar{q}}}
\bigtriangledown^2V_{\mathrm{coul}},
\end{equation}
with the last term being the hyperfine interaction that causes the
mass splitting between vector and pseudoscalar mesons. Here we
choose the confining potential (the second term) to be the harmonic
oscillator potential rather than the linear potential in order to
keep consistency with our harmonic oscillator wave function for the
quarkonium. The values of parameters $a$, $b$ and $\alpha_s$ were
given in Ref~\cite{Choi99}.

The mass of the meson is obtained as $M_{q\bar{q}}= \langle \phi
|H_{q\bar{q}}| \phi \rangle$~\cite{Choi99}. For the 2S quarkonium,
we have
\begin{eqnarray}
M_{q\bar{q}}&=&\frac{16}{3\sqrt{\pi}\beta^7}\int_0^\infty
(\sqrt{m^2_q+p^2})p^2(p^2-\frac{3}{2}\beta^2)^2e^{-p^2/
\beta^2}\mathrm{d}p\nonumber\\
&&+a+\frac{7b}{2\beta^2}-\frac{20\alpha_s\beta}{9\sqrt{\pi}}\nonumber\\
&&+\left\{\begin{array}{l} \frac{4\alpha_s\beta^3}{3\sqrt{\pi}m_q^2}
~~\mathrm{(vector~~quarkonia)},\\
-\frac{4\alpha_s\beta^3}{\sqrt{\pi}m_q^2}
~~\mathrm{(pseudoscalar~~quarkonia)},
\end{array}\right.
\end{eqnarray}
and such formula of the mass of the 1S quarkonium can be found in
Ref~\cite{Choi09}.

\section{\label{sec:level3} Numerical results}

In our numerical calculation, the parameter $\beta$ in the wave
function and the mass of the constituent quark $m_q$ were chosen to
fit experimental data. Since the only difference between the vector
and pseudoscalar quarkonia that share the same energy quantum
number~($n$) is the hyperfine interaction term in this model, we
choose the same $\beta$ for them. For charmonia, $m_c$ and
$\beta_{J/\psi}~(\beta_{\eta_c})$ were fixed by
Refs.~\cite{Choi07,Choi09}, and their results are in good agreement
with experimental data, so we use their values of $m_c$ and
$\beta_{J/\psi}~(\beta_{\eta_c})$, and we only have to fix the
parameter $\beta_{\psi'}(\beta_{\eta_c'})$. For the bottomoina, we
fix all the parameters $m_b$, $\beta_{\Upsilon}(\beta_{\eta_b})$ and
$\beta_{\Upsilon'}(\beta_{\eta_b'})$.

The parameters of the charmonia are fixed as
\begin{eqnarray}
&&m_c=1.8~\mathrm{GeV},~~\beta_{J/\psi}(\beta_{\eta_c})
=0.6998~\mathrm{GeV},\nonumber\\
&&\beta_{\psi'}(\beta_{\eta_c'})=0.630~\mathrm{GeV},
\end{eqnarray}
and our numerical results of 2S charmonia are listed in
Table~\ref{charmonia}.
\begin{table}[h]
\caption{\label{charmonia} Numerical results (GeV) of 2S charmonia.}
\begin{tabular}{lcc}
\hline \hline
& Experiment~\cite{PDG10} & Theory\\
\hline $F_{\psi' \rightarrow\eta_c \gamma^* }(0)$
& 0.0392$\pm$0.0031 & 0.0402 \\
$f_{\psi'}(\psi' \rightarrow e^+ e^-)$
& 0.1910$\pm$0.0057 & 0.2474 \\
$M_{\psi'}$ & 3.686093$\pm$0.000034 & 3.778 \\
$M_{\eta_c'}$ & 3.637$\pm$0.004 & 3.637\\
$F_{\psi' \rightarrow\eta_c' \gamma^* }(0)$
& $<0.9006$ & $0.6292$ \\
$F_{\eta_c' \rightarrow\gamma\gamma^*}(0)$
& $<0.0590$ & $0.0271$ \\
\hline \hline
\end{tabular}
\end{table}
The numerical results of 1S charmonia can be found in
Refs.~\cite{Choi07,Choi09}. The transition form factors $F_{\psi'
\rightarrow\eta_c' \gamma^* }(0)$ and $F_{\eta_c'
\rightarrow\gamma\gamma^*}(0)$ are our predictions, and we see from
the table that they are well below experimental upper limits. We can
also obtain the branching ratios of the two decay modes using
Eqs.~(\ref{decay width a}) and (\ref{decay width b}) and the total
widths of $\psi'$ and $\eta_c'$~\cite{PDG10}:
\begin{eqnarray}
\mathrm{Br}(\psi' \to \gamma
\eta_c')&=&3.9012\times10^{-4},\nonumber\\
\mathrm{Br}(\eta_c' \rightarrow 2\gamma)&=&1.0555\times10^{-4}.
\label{brcharmonia}
\end{eqnarray}

BES-III collaboration reported very recently the first measurement
of the branching ratio Br$(\psi'\to\gamma\eta_c')= (4.7\pm
0.9_{\mathrm{stat}}\pm 3.0_{\mathrm{sys}})\times 10^{-4}$
~\cite{BES11}, and our prediction in Eq.~(\ref{brcharmonia}) is in
agreement with the preliminary data within error bars. The very
recent theoretical study in Ref.~\cite{Li11} gives
$\Gamma(\psi'\to\gamma\eta_c')=0.08^{+0.03} _{-0.03}~\mathrm{keV}$,
and converted into the branching ratio using the total width of
$\psi'$~\cite{PDG10}, it is Br$(\psi'\to\gamma\eta_c')=(2.7972\pm
1.1)\times 10^{-4}$. Other theoretical predictions for
Br$(\psi'\to\gamma\eta_c')$ fall in a range of $(0.1-6.2)\times
10^{-4}$~\cite{CLEO10}.

Assuming that $\eta_c$ and $\eta_c'$ have equal branching fractions
to $K_SK\pi$, Ref.~\cite{CLEO04} obtained the experimental data
$\Gamma_{\gamma\gamma}(\eta_c')=1.3\pm0.6~\mathrm{keV}$, using the
total width of $\eta_c'$~\cite{PDG10}, the branching ratio is
$\mathrm{Br} (\eta_c' \rightarrow 2\gamma)=(0.9286\pm0.63)\times
10^{-4}$. The theoretical prediction in Ref.~\cite{Chao96},
converted into branching ratio, gives $\mathrm{Br}(\eta_c'
\rightarrow 2\gamma)=1.4286\times 10^{-4}$. We see that our
prediction in Eq.~(\ref{brcharmonia}) is close to these data.
However, the accurate experimental data for this decay mode is still
not available, and only the upper limit $\mathrm{Br} (\eta_c'
\rightarrow 2\gamma)<5\times10^{-4}$ is given~\cite{PDG10}. Future
experimental measurements at BES and CLEO may provide tests for
these predictions of Br$(\psi'\to\gamma\eta_c')$ and
$\mathrm{Br}(\eta_c' \rightarrow 2\gamma)$.

For bottomonia, the parameters are fixed as
\begin{eqnarray}
&&m_b=5.1~\mathrm{GeV},~~\beta_{\Upsilon}(\beta_{\eta_b})
=1.1656~\mathrm{GeV},\nonumber\\
&&\beta_{\Upsilon'}(\beta_{\eta_b'})=1.1050~\mathrm{GeV},
\end{eqnarray}
and our numerical results of the 1S and 2S bottomonia are listed in
Table~\ref{bottomonia}.
\begin{table}[h]
\caption{\label{bottomonia} Numerical results (in units of GeV) of
the 1S and 2S bottomonia.}
\begin{tabular}{lcc}
\hline \hline
& Experiment~\cite{PDG10} & Theory\\
\hline $f_{\Upsilon'}(\Upsilon' \rightarrow e^+ e^-)$
& 0.1657$\pm$0.0097 & 0.1944\\
$f_{\Upsilon}(\Upsilon \rightarrow e^+ e^-)$
& 0.2384$\pm$0.0044 & 0.1822\\
$F_{\Upsilon' \rightarrow\eta_b \gamma^*}(0)$
& -0.0047$\pm$0.0009 & -0.0047\\
$M_{\Upsilon'}$ & 10.0233$\pm$0.00031 & 10.0592\\
$M_{\Upsilon}$ & 9.46030$\pm$0.00026 & 9.4087\\
$M_{\eta_b}$ & 9.3909$\pm$0.0028 & 9.3346\\
$M_{\eta_b'}$ & ? & 9.9644\\
$F_{\eta_b' \rightarrow\gamma\gamma^*}(0)$ & ? & 0.0019\\
$F_{\eta_b \rightarrow\gamma\gamma^*}(0)$ & ? & 0.0020\\
$F_{\Upsilon' \rightarrow\eta_b' \gamma^*}(0)$ & ? & -0.1216\\
$F_{\Upsilon \rightarrow\eta_b \gamma^*}(0)$ & ? & -0.1261\\
\hline \hline
\end{tabular}
\end{table}
Our results give the prediction $\Gamma(\eta_b'\to
2\gamma)=0.1494~\mathrm{keV}$, compared with the prediction given in
Ref.~\cite{Chao96} $\Gamma(\eta_b'\to 2 \gamma)=0.21~\mathrm{keV}$.
Future experiments at LHC or by the Belle experiment on $\eta_b'$
can not only test these predictions, but also help us learn more
about this meson by providing more experimental information about
it.

The small values of the experimental data for the branching ratio of
the mode $V(2S)\to \gamma P(1S)$ in Table~\ref{charmonia} and
Table~\ref{bottomonia} can be easily understood with our wave
functions, as the $2S$ and $1S$ wave functions are orthogonal to
each other and their overlap in Eq.~(\ref{fvp}) is suppressed.

Although nonrelativistic models have provided efficient and powerful
theoretical tools to handle various problems related to hadron
structure, the relativistic models have been successful in many
investigation of hadron
structures~\cite{Xiaoetal1,Xiaoetal2,Xiaoetal3,Xiaoetal4,Xiaoetal5,Ma93,MaMeson1,MaMeson2,MaNucleon1,MaNucleon2,MaNucleon3,MaNucleon4,MaNucleon5,MaNucleon6,MaNucleon7}.
It is thus necessary to make an estimate of the effect due to
nonrelativistic to relativistic treatments~\cite{Hwang}. For
simplicity, we assess the non-relativistic to relativistic effects
by letting $\mathbf{k}_\perp^2$ of $\sqrt{\mathbf{k}_\perp^2+m_q^2}$
in the expressions of the decay constants and form factors to be
zero. For examples, after this procedure, we have $F_{\psi'
\rightarrow \eta_c \gamma^* }(0)=0.1167~\mathrm{GeV}$, compared with
$F_{\psi'
\rightarrow\eta_c\gamma^*}(0)=0.0402~\mathrm{GeV}/0.0392~\mathrm{GeV}$
from the relativistic treatment / experimental data, and
$f_{\Upsilon'}(\Upsilon' \rightarrow e^+ e^-)=0.2159~\mathrm{GeV}$
compared with $f_{\Upsilon'}(\Upsilon' \rightarrow e^+
e^-)=0.1944~\mathrm{GeV}/0.1657~\mathrm{GeV}$ from the relativistic
treatment / experimental data. We see that the relativistic
treatment is needed to describe the experimental data well.

\section{\label{sec:level4} summary}

In this work, we studied the 2S quarkonia $\psi'$, $\eta_c'$,
$\Upsilon'$ and $\eta_b'$ in light-front quark model. Similar with
the 1S harmonic oscillator wave function that was commonly used as
the 1S quarkonium wave function in light-front quark model studies,
we tried to use the 2S harmonic oscillator wave function as the 2S
quarkonium wave function. The decay constants and transition form
factors of these quarkonia are calculated. Using the QCD-motivated
Hamiltonian for mesons, we also calculated masses of these
quarkonia. Our numerical results of these quantities are in
agreement with experimental data. Predictions of transition form
factors and masses of these quarkonia are made, and these
predictions can be tested by future experiments. The 1S and 2S wave
functions could also be used as inputs to study other problems such
as the ``$\rho \pi$ puzzle" and the suppression of heavy-quarkonia
at RHIC energy.


 This work is partially supported by National Natural
Science Foundation of China (Grants No.~11021092, No.~10975003,
No.~11035003, and No.~11120101004) and by the Research Fund for the
Doctoral Program of Higher Education (China).

%
%

\end{document}